\documentclass[prb,twocolumn,preprintnumbers,amsmath,aps]{revtex4}

\usepackage{amssymb,amsfonts,amsmath}
\usepackage{graphicx}

\begin{document}

\title{Avalanches and dimensional reduction breakdown in the critical behavior of disordered systems}

\author{Gilles Tarjus} \email{tarjus@lptl.jussieu.fr}
\affiliation{LPTMC, CNRS-UMR 7600, Universit\'e Pierre et Marie Curie,
bo\^ite 121, 4 Place Jussieu, 75252 Paris c\'edex 05, France}

\author{Maxime Baczyk} \email{tissier@lptl.jussieu.fr}
\affiliation{LPTMC, CNRS-UMR 7600, Universit\'e Pierre et Marie Curie,
bo\^ite 121, 4 Place Jussieu, 75252 Paris c\'edex 05, France}
\author{Matthieu Tissier} \email{tissier@lptl.jussieu.fr}
\affiliation{LPTMC, CNRS-UMR 7600, Universit\'e Pierre et Marie Curie,
bo\^ite 121, 4 Place Jussieu, 75252 Paris c\'edex 05, France}

  \begin{abstract} 
  We investigate the connection between a  formal property of the critical behavior
of several systems in the presence of quenched disorder, known as ``dimensional reduction'', and the presence in the
same systems at zero temperature of collective events known as ``avalanches''. Avalanches generically produce nonanalyticities
in the functional dependence of the cumulants of the renormalized disorder. We show that this leads to a
breakdown of the dimensional reduction predictions if and only if  the fractal dimension characterizing the scaling properties of
the avalanches is exactly equal to the difference between the dimension of space and the scaling dimension of the primary field,
\textit{e.g.} the magnetization in a random field model. This is proven by combining scaling theory and functional renormalization
group. We therefore clarify the puzzle of why dimensional reduction remains valid in random field systems above a nontrivial
dimension (but fails below), always applies to the statistics of branched polymer and is always wrong in elastic models of interfaces
in a random environment. 
\end{abstract}

\maketitle

In the theory of disordered systems, ``dimensional reduction'' is the property shared by some models that 
the long-distance physics in the presence of quenched disorder in some spatial dimension $d$ is the same 
as that of the pure model with no disorder in a reduced spatial dimension, $d-2$. In the known examples 
where it has been found through perturbation theory, \textit{i.e.} the random field Ising model (RFIM)~\cite{aharony76,grinstein76,young77}, 
elastic manifolds in a random environment~\cite{efetov77}, the random field and random anisotropy O(N) models~\cite{aharony76,grinstein76,young77,fisher85} and the statistics of dilute  branched polymers\footnote[1]{We include 
the branched polymer problem in the list of disordered systems by an abuse of language. Indeed, there is no quenched disorder 
in this case but the equivalence comes from the $n \rightarrow 0$ limit which is common to the field-theoretical 
description of self-avoiding polymer chains and to the replica theory of disordered systems. The 
dimensional reduction then leads to the Yang-Lee edge singularity in dimension $d-2$. See Ref.~\cite{parisi81}.}, 
it entails two conditions: (1) that the long-distance physics is controlled by a zero-temperature fixed point, so that it 
can be equally described from the solution(s) of a stochastic field equation at zero temperature, and (2) that an underlying 
supersymmetry (involving rotational invariance in superspace)  emerges in the field-theoretical treatment of the stochastic equation~\cite{parisi79,parisi81,wiese05}. Dimensional reduction however is known to be wrong in some of the above models, 
the random field and random anisotropy models in low enough dimension (a rigorous proof exists for the RFIM in  $d=2$ and $3$~\cite{imbrie84,bricmont87}) and the elastic manifold in a random environment~\cite{efetov77}. On the other hand, it is proven 
to be right for the branched polymer case in all dimensions below the upper critical one~\cite{brydges03,cardy03}. 

We have recently shown that the breakdown of dimensional reduction and the spontaneous breaking of the underlying 
supersymmetry take place below a nontrivial critical dimension\footnote[2]{Actually, the two phenomena take place at 
two very close but distinct critical dimensions $d_{DR}$ and $d_{cusp}$, with  $d_{DR}\lesssim d_{cusp}$; the former 
represents the point at which supersymmetry is spontaneously broken along the flow and the dimensional-reduction fixed 
point vanishes and the latter is where the dimensional-reduction fixed point becomes unstable to a nonanalytic perturbation. 
For the RFIM, the two are numerically almost indistinguishable and for the RFO(N)M  one finds $N_{DR}=18$ and 
$N_{cusp}=18.393..$ when approaching $d=4$ [G. Tarjus, M. Baczyk, and M. Tissier, in preparation (2012)].} in the 
random-field Ising, and more generally O(N), model: this dimension is close to $5$ for the Ising ($N=1$) version and 
decreases continuously as $N$ increases until it reaches $4$ when $N$ approaches $18$ (the upper critical dimension is 
equal to $6$ for random field systems)~\cite{tarjus04,tissier06,tissier11}.  Describing this phenomenon requires a  renormalization 
group (RG) approach that is \textit{functional}, as the origin of the dimensional reduction breakdown is the appearance of a 
nonanalytic dependence of the renormalized cumulants of the random field (a linear ``cusp'') in the dimensionless fields, and 
\textit{nonperturbative}, as it takes place away from regimes where some form of perturbation analysis is possible 
(except for the O(N) model when $d$ is close to the lower critical dimension of 4~\cite{tarjus04,tissier11}). A similar conclusion 
was previously reached for the case of an elastic manifold in a random environment, but in this model the dimensional reduction 
predictions fail for all dimensions at and below the upper critical dimension (here equal to $4$) and can be already assessed through 
a functional but perturbative RG~\cite{fisher86b,narayan92,FRGledoussal-chauve,FRGledoussal-wiese}. What is the physical 
mechanism behind these seemingly formal results?

The existence of a cusp in the cumulants of the renormalized disorder can be assigned to the presence of collective events known 
as ``avalanches''. In any typical sample of a disordered model, the ground state, which is the relevant configuration that describes 
the equilibrium properties of the system at zero temperature, abruptly changes for specific values of the external source; the location 
of these abrupt changes are sample-dependent and the configurational change between two ground states is precisely an 
avalanche~\cite{static_middleton,static-distrib_ledoussal,frontera00,dukowski03,wu05,liu09,monthus11}. (The latter is sometimes 
called a ``static'' avalanche\footnote[3]{In the context of elastic interfaces in a disordered environment, the static avalanches are also 
called ``shocks'' by analogy with  the behavior found in the Burgers equation~\cite{BBM96,ledoussal06}.}.) The same phenomenon 
is observed, still at zero temperature, when the system is driven by the external source without being allowed to equilibrate. 
The corresponding ``dynamic'' avalanches then take place out of equilibrium, between two metastable states of the system~\cite{dynamic_rosso,zapperi98,perkovic99,sethna01,perez04,liu09}.

However, the fact that abrupt changes corresponding to discontinuous variations of the magnetization (to use the language of magnetic 
systems) are found at zero temperature should come as no surprise. In disordered systems, this can take place even in noninteracting 
zero-dimensional models. Consider for instance a $d=0$ (single point) $\phi^4$ theory with parameters such that the potential has two 
minima and couple the field $\phi$ to a random source $h$ and a controllable source $J$. Then, according to the value of $h+J$, the ground 
state of the system will switch from the vicinity of one minimum to that of the other one with a jump in the magnetization (the field $\phi$) 
when $h+J=0$. This jump corresponds to an avalanche, albeit a zero-dimensional one. As will be illustrated in more detail below, these 
avalanches do generate a cusp in the cumulants of the effective random field (see Fig.~1). However,  avalanches, and the resulting 
nonanalyticities in renormalized disorder cumulants,  affect the long-distance physics of a $d$-dimensional disordered model and play 
a role in the failure of dimensional reduction only if they are of collective origin and occur on sufficiently large scales.

From the above discussion we conclude that in disordered systems, (i) the physics of dimensional reduction and its breakdown 
is associated with an effective zero-temperature theory, (ii) avalanches corresponding to discontinuous jumps of the magnetization 
and more generally of the field configuration are the rule rather than the exception at zero temperature and (iii) the dimensional-reduction 
breakdown results from the presence of  avalanches that must be collective enough to influence the long-distance properties of the 
model. The central question is then: Under which conditions does this happen? We show that dimensional reduction remains valid 
when the exponent $d_f$ that characterizes the scaling behavior of the largest typical  avalanches at criticality is strictly less than the difference between the spatial dimension $d$ and the scaling dimension of the field $d_{\phi}$ near the relevant 
zero-temperature fixed point: $d_f < d- d_{\phi}$. This condition is satisfied for the RFIM at and close to the upper critical dimension $d_{uc}=6$  and for branched polymers at and around the upper critical dimension $d_{uc}=8$ and below $d=4$ (at least). 
On the other hand, dimensional reduction breakdown takes place when $d_f$ is equal to $d- d_{\phi}$\footnote[4]{The condition 
$d_f > d- d_{\phi}$ leads to unphysical results, as discussed further down}. This is found for elastic manifolds in a random 
environment at and near the upper critical dimension $d_{u}=4$ as well as for the RFIM below a critical dimension close to $5$. 
This clarifies the intriguing result of why dimensional reduction fails below a nontrivial dimension for the RFIM, is always 
broken for random elastic manifolds, but applies to the branched-polymer problem.

\textbf{Avalanches and their consequences on the cumulants of the renormalized disorder.}
Take a disordered system at zero temperature in which avalanches, representing discontinuous changes of the relevant configuration 
of the system under a variation of an external source, are present. We use the language of magnetic systems and characterize 
configurations by the local magnetization.  All considerations, however, equally well apply to configurations described by a 
continuous field, in or out of equilibrium, and to nonmagnetic systems, \textit{e.g.} to the displacement of an interface in a 
random medium~\cite{static_middleton,static-distrib_ledoussal,dynamic_rosso,dynamic_av-distrib_ledoussal} or to the appropriate 
density field describing the universal properties of dilute branched polymers. We focus on situations in which the field under study 
(the ``magnetization'') is the local order parameter and is linearly coupled to the external source, but one should keep in mind that 
avalanches could also be triggered by changing for instance the strength of the disorder~\cite{bray87,RFIM_chaos_alava} at 
zero temperature and that avalanches may involve a field that is not the local order parameter (\textit{e.g.}, the magnetization in a spin glass~\cite{SKav_young,SKav_pazmandi,SKav-distrib_ledoussal}). 

Consider for simplicity an external source $J$ that is uniform in space. The avalanches can then be characterized by their size 
$S$ (the overall change in the \textit{total}
magnetization\footnote[5]{In principle, the size of an avalanche, if
  defined in terms of the total number of spins that flip in an Ising
  model or of an equivalent geometric quantity in a field theory,
  could be different from the associated change in magnetization. This
  is for instance what is predicted for spin glasses by the droplet
  theory~\cite{droplet_fisher}. However, for the systems under study, one expects and proves in several cases that the size and the magnetization are essentially the same and scale in the same way with the (linear) spatial extent of the avalanche.}) whose distribution is described by a density $\rho(S, J)$, such that  
$\rho(S, J)dS\, dJ$ is the (disorder averaged) number of avalanches of size between $S$ and $S+dS$ when the source is between 
$J$ and $J+dJ$. The magnetization $m(J;\mathbf h)$ is the spatial average of the local order parameter field for a given sample 
characterized by the disorder realization $\mathbf{h}$. Its change between two values of the  external source $J_1$ and $J_2$ is 
the sum of two contributions: a first one comes from the smooth changes in the ground state (in equilibrium) or in the metastable 
state (out of equilibrium) and another one comes from the avalanches that take place between $J_1$ and $J_2$ (with $J_2 > J_1$). 
As a consequence, the moments of the difference $[m(J_1;\mathbf h)-m(J_2;\mathbf h)]$, which is a random variable, are given by
\begin{equation}
\begin{aligned}
\label{eq_moments_aval}
&\overline{[m(J_2;\mathbf h)-m(J_1;\mathbf h)]^p}=\\& (\frac{1}{L^d})^p\int_{J_1}^{J_2}dJ' \int_{S_{min}}^{\infty} 
dS \;S^p \rho(S,J')+ reg\;,
\end{aligned}
\end{equation}
where the first term is due to avalanches taking place at the same value of the external source and the second term, denoted 
by $reg$, includes the contributions involving the smooth variation of the magnetization or distinct avalanches. $S_{min}$ is 
a microscopic lower cutoff on the size of the avalanches, $L^d$ is the sample volume, and the overline denotes the average 
over the quenched disorder.

For even moments, due to the symmetry in the exchange of $J_1$ and $J_2$, the first term in Eq.~(\ref{eq_moments_aval}) gives 
rise to a linear cusp when $J_2 \rightarrow J_1$:
\begin{equation}
\begin{aligned}
\label{eq_cusp1}
&\overline{[m(J_2;\mathbf h)-m(J_1;\mathbf h)]^{2p}}\\&=\vert J_2-J_1\vert L^{-2p d} \int_{S_{min}}^{\infty} dS\; 
S^{2p} \rho(S,J) + O([J_2-J_1]^2).
\end{aligned}
\end{equation}
[The other term in Eq.~(\ref{eq_moments_aval}) is regular or at least less singular than the first one and is therefore a $O([J_2-J_1]^2)$.] 

It is easily realized that the $p$th moment is obtained by considering disorder averages over $p$ copies of the \textit{same} 
sample, with each copy coupled to a distinct external source $J_a$, $a=1,\cdots,p$. The first moment is the change in the average 
magnetization of the system and the $p$th cumulant (which is the connected piece of the $p$th moment) is then related to Green's 
functions at zero momentum of a $p$-copy system in which each copy is characterized by a distinct source. For example, the 
relevant 2-point Green's function, given by
$\widetilde{G}(q=0;J_1,J_2) =L^d [\overline{m(J_1;\mathbf h) m(J_2;\mathbf h)}-\overline{m(J_1;\mathbf h)}\;
\overline{m(J_2;\mathbf h)}]$, is an extension to general sources $J_1 \neq J_2$ of what is usually called the ``disconnected'' 
2-point correlation function in the theory of disordered systems. One-particle irreducible (1PI) correlation functions 
(or proper vertices)~\cite{zinn-justin89} associated with the above Green's functions can be introduced along the same 
lines\footnote[6]{For the equilibrium case, this is simply done through a Legendre transformation that allows one to go 
from the generating functional depending on the external sources to the one  depending on the local magnetization (the 
Gibbs free-energy in the language of magnetic systems) or the classical field (the effective action in the language of 
quantum field theory). This can be generalized to the out-of-equilibrium case but requires the introduction of auxiliary fields.}. 
From Eq.~(\ref{eq_cusp1}), one immediately derives that, for instance, the $2$-point Green's function 
$\widetilde G(q=0; J_1=J-\delta J, J_2=J+\delta J)$ has a nonanalytic dependence as $\delta J \rightarrow 0$, with the 
amplitude of the cusp related to the second moment of the avalanches:
\begin{equation}
\label{eq_cusp_d}
\begin{split}
&\widetilde{G}(0;J-\delta J,J+\delta J) - \widetilde{G}(0;J,J)  =\\& -  \vert \delta J\vert \frac{1}{L^{d}} 
\int_{S_{min}}^{\infty} dS\; S^2 \rho(S,J)+ O(\delta J^2).
\end{split}
\end{equation}
This can be transposed to the associated 1PI vertices and can generalized to higher orders as well\footnote[7]{We have 
discussed the case where the relevant variable or field involved in the avalanches has a single component. Cusps and 
avalanches are also present  in the case of multi-component fields (see \textit{e.g.}, Ref.~\cite{dasilveira99}). However, 
as the analysis is more complex, we mostly restrict ourselves here to the consideration of systems with a single-component field.}.

\begin{figure}[ht]
\begin{center}
(a)\includegraphics[width=.5 \linewidth]{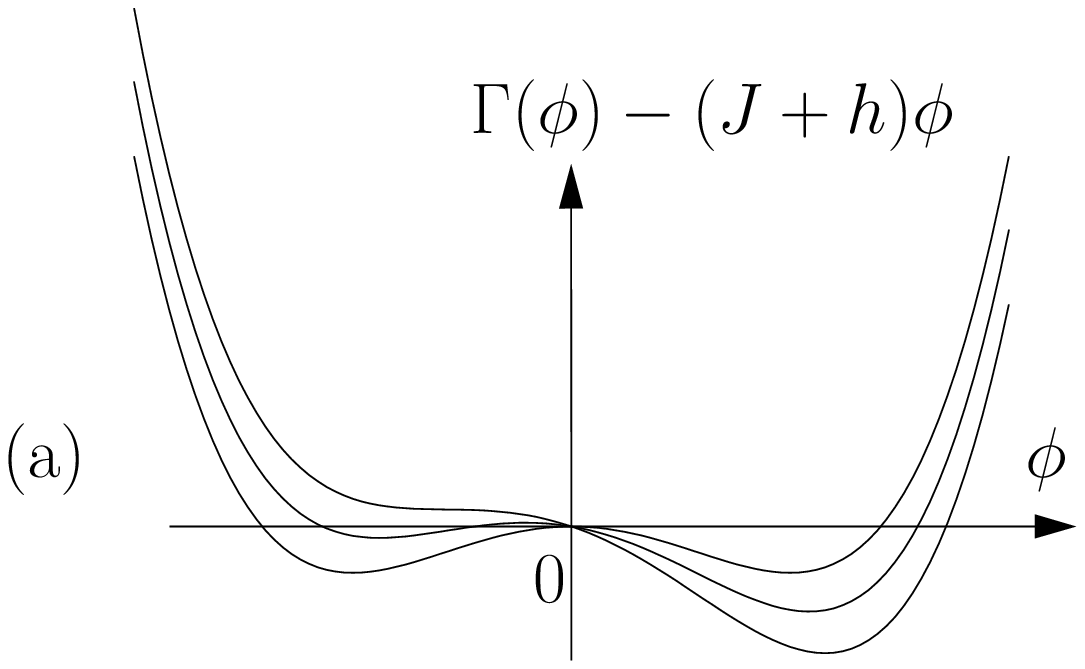}
\end{center}

\begin{center}
(b) \includegraphics[width=.5 \linewidth]{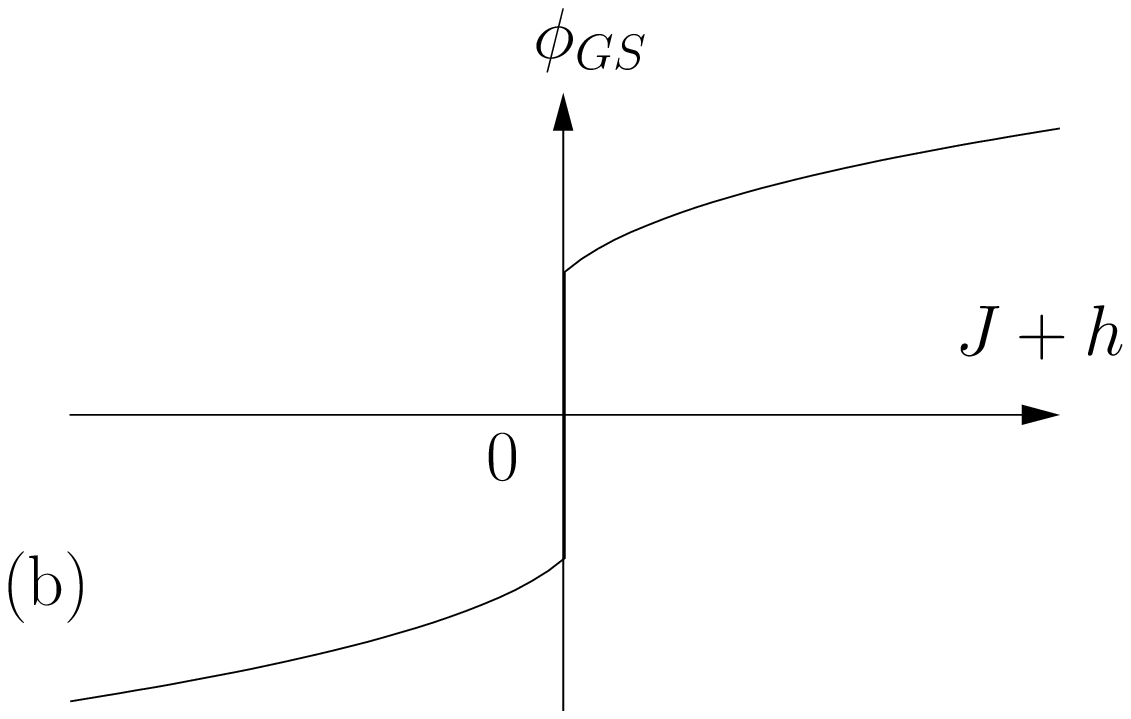}
\end{center}

\begin{center}
(c)\includegraphics[width=.5 \linewidth]{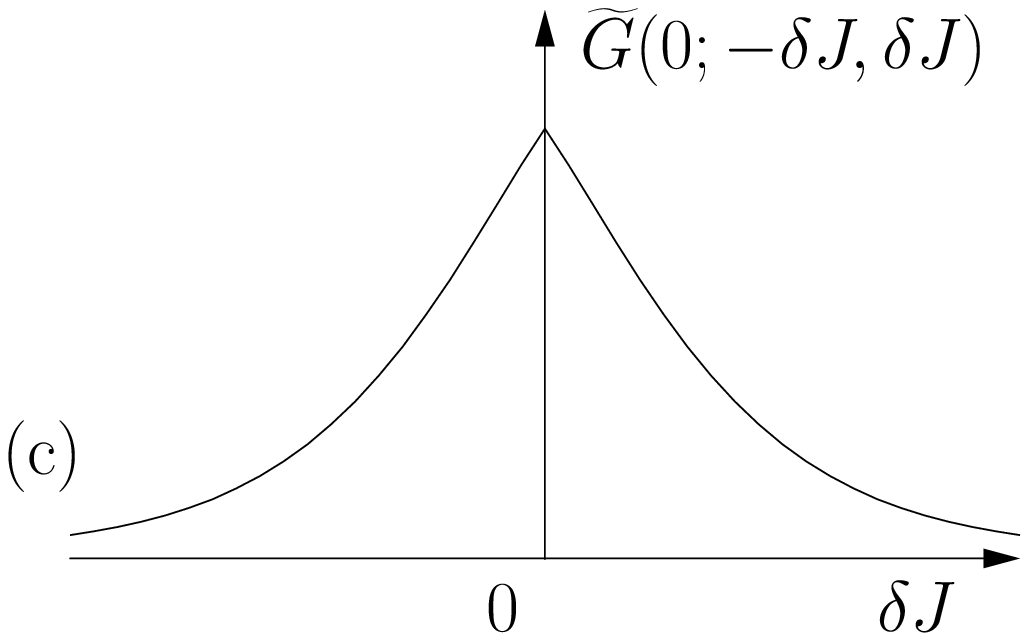}
\end{center}

\caption{ Illustration of avalanches and their consequence on the functional dependence of the Green's functions in the 
schematic case of the $d=0$ RFIM studied in equilibrium at $T=0$. (a) Free energy $\Gamma(\phi)-(J+h)\phi$ versus 
$\phi$ for different values of $J$, with  $\Gamma(\phi)=-(\vert \tau \vert/2)\phi^2+(g/4!) \phi^4$; (b) Ground state configuration 
associated with (a); Two-point Green's function $\widetilde{G}(0;-\delta J,\delta J) = [\overline{\phi_{GS}(-\delta J+ h)
\phi_{GS}(\delta J+ h)}-\overline{\phi_{GS}(-\delta J+ h)}\;\overline{\phi_{GS}(\delta J+ h)}]$, where the average is over a 
Gaussian distributed random field $h$: notice the linear cusp around $\delta J=0$. }
\end{figure}

The above arguments show that avalanches induce a linear cusp in the functional dependence of the correlation functions 
describing the cumulants of the effective or renormalized disorder at $T=0$. This is always true even when the avalanches 
take place on a restricted scale, away from any critical conditions (see Fig.~1 for an illustration). However, we are interested 
in the long-distance behavior of disordered systems. We therefore study the situation in which avalanches occur on all scales, 
as found for instance in the RFIM at the equilibrium or out-of-equilibrium critical point, in the rough phase of an elastic 
manifold pinned by a random medium or at its depinning transition, etc.

Consider a $d$-dimensional system of linear size $L$ at zero temperature. At large scale, when the correlation length and 
the extent of the largest typical avalanches have reached the system size, one expects that the (normalized) probability density 
of having an avalanche of size $S$ (see footnote 5) can be written in the following scaling form~\cite{dahmen96,perez04,liu09,monthus11}:
\begin{equation}
\label{eq_distrib_avalanche_L}
D_L(S,J)=S^{-\tau}\; \mathcal D(\frac{S}{S_L}, \vert J-J_c\vert S^{\psi})
\end{equation}
where $S_L\sim L^{d_f}$ is the size the largest typical  ``critical'' avalanches in the finite system\footnote[8]{In a finite-size 
system it may be important to appropriately sort out the various classes of large, system-spanning avalanches: the relevant 
ones for the critical scaling have been called ``spanning critical''  in the context of the out-of-equilibrium RFIM~\cite{perez04} 
and $d_f$ is then the associated exponent.}; $S_L$ acts as a cutoff for the scaling function $\mathcal D$ that decays exponentially 
for $S/S_L \gtrsim 1$. Critical conditions correspond to $J=J_c$ (for the RFIM at equilibrium one has $J_c=0$ due to the $Z_2$ 
symmetry and for the random manifolds there is no condition on $J$ as the whole phase is critical). The avalanche size distribution 
is normalized so that $\int_{S_{min}}^{\infty} dS \;D_L(S,J) =1$ and the moments of the normalized avalanche size distribution 
are then defined as $<S^p>_L=\int_{S_{min}}^{\infty} dS \; S^p D_L(S,J)$. When $1<\tau<2$, which is usually found (for 
instance, the mean-field value for $\tau$ is equal to $3/2$ in all models), the normalization factor is dominated by the small 
avalanches whereas all moments with $p\geq 1$ are dominated by the largest typical avalanches and behave as $(S_L)^{p+1-\tau}
\sim L^{(p+1-\tau)d_f}$ when $L\rightarrow \infty$.

We keep using the language of magnetic systems and let $m_L(J;\mathbf h)$ denote the magnetization of a given sample of 
linear size $L$. (Here and below, we explicitly indicate the dependence on the system size $L$; this makes the expressions 
somewhat clumsy but will be helpful later on to make the connection with the results of the  functional RG.) We are primarily 
interested in the moments of the random variable $m_L(J_2;\mathbf h)-m_L(J_1;\mathbf h)$. The density of avalanches [see Eq.~(\ref{eq_moments_aval})] is related to the normalized probability density by an overall $L$ and $J$ dependent factor: 
$\rho_L(S,J)=\rho_{0,L}(J)D_L(S,J)$. As discussed before in connection to Eq.~(\ref{eq_moments_aval}), the first moment 
of $m_L(J_2;\mathbf h)-m_L(J_1;\mathbf h)$ contains a contribution from the smooth change of the magnetization and one 
from the avalanches. The so-called  ``connected'' susceptibility $\chi_{c,L}(J)$, which is the standard magnetic susceptibility 
divided by the temperature in order to have a proper zero-temperature limit and which is obtained by deriving $m_L$ with 
respect to $J$, can then be expressed as
\begin{equation}
\begin{aligned}
\label{eq_chi_cL}
\chi_{c,L}(J)=\chi_{c,_L}^{smooth}(J) + \frac{1}{L^d} \int_{S_{min}}^{\infty} dS\, S \,\rho_L(S,J).
\end{aligned}
\end{equation}
Under critical conditions, $\chi_{c,L}$ goes as $L^{2-\eta}$. By further making the natural assumption that the contribution 
from the avalanches is of the order or larger than the smooth one and by using Eqs.~(\ref{eq_distrib_avalanche_L},\ref{eq_chi_cL}) 
as well as the fact that the first moment of the avalanches $<S>_L$ is dominated by large avalanches, one then obtains that
\begin{equation}
\label{eq_scaling_chi_cLmore}
\rho_{0,L}(J_c) L^{-d+(2-\tau)d_f} \sim L^{2-\eta}.
\end{equation}
As a result, Eq.~(\ref{eq_cusp1}) can be rewritten as
\begin{equation}
\begin{aligned}
\label{eq_cusp_p}
&\overline{[m_L(J_2;\mathbf h)-m_L(J_1;\mathbf h)]^p}\\& \sim \vert J_2-J_1\vert \, L^{2-\eta -(p-1)(d-d_f)}+ O((J_2-J_1)^2)
\end{aligned}
\end{equation}
and the linear cusp in the 2-point Green's function at zero momentum $\widetilde{G}_L(q=0;J-\delta J, J+\delta J)$ when 
$\delta J \rightarrow 0$ [see eq.~(\ref{eq_cusp_d})] is found as
\begin{equation}
\begin{aligned}
\label{eq_greenfunction_cusp_crit}
&\widetilde{G}_L(q=0;J_c-\delta J,J_c+\delta J)-\widetilde{G}_L(q=0;J_c,J_c)\\& \sim  \vert \delta J\vert\; L^{d_f +2-\eta}
\end{aligned}
\end{equation}
up to a $O(\delta J^2)$. The amplitude of the cusp therefore diverges as the size of the system diverges. (One should however 
keep in mind that the whole function $\widetilde{G}_L(q=0)$ itself diverges as $L^{4-\bar \eta}$ at criticality.) The above result 
generalizes to higher-order Green's functions through their relation to the cumulants of the magnetization.

From Eq.~(\ref{eq_greenfunction_cusp_crit}), it is easily derived that the associated 1PI correlation function $\Delta_L(q=0;m_1,m_2)$, 
which is the second cumulant of the renormalized disorder, also has a cusp in $\vert m_2-m_1\vert$ as $m_2\rightarrow m_1$. After 
introducing $m_1=m_c -\delta m$ and $m_2=m_c+\delta m$, where  $\delta m \rightarrow 0$ and $m_c$ corresponds to 
the value of the average magnetization at criticality, and using the relation between Green's functions and 1PI functions 
(see ref. \cite{zinn-justin89} and footnote 6) as well as $\delta m=\overline{m_L(J_c+\delta J;\mathbf h)}- 
\overline{m_L(J_c-\delta J;\mathbf h)}\simeq \delta J\, \chi_{c,L}(J_c)$ when $\delta J \rightarrow 0$, we obtain, 
up to a $O(\delta m^2)$,
\begin{equation}
\begin{aligned}
\label{eq_1PIfunction_cusp}
&\Delta_L(q=0;m_c-\delta m,m_c+\delta m)-\Delta_L(q=0;m_c,m_c) \simeq \\& \chi_{c,L}(J_c)^{-2}
[\widetilde{G}_L(q=0;J_c-\delta J,J_c+\delta J)-\widetilde{G}_L(q=0;J_c,J_c)]\\&
\sim \vert \delta m\vert\; L^{d_f -2(2-\eta)}
\end{aligned}
\end{equation}
where we have also used that $\chi_{c,L}(J_c)\sim L^{2-\eta}$. Again, the above expression can be generalized to higher-order 
cumulants of the renormalized disorder, which all display a cusp in their functional dependence with an amplitude that is system-size 
dependent at criticality.

\textbf{Functional RG and dimensional reduction breakdown.} 
As already stressed, the functional RG is a powerful and necessary framework to describe the critical behavior of the disordered systems 
of interest. Within such an approach, which is a version of Wilson's continuous RG~\cite{wilson74,wegner73,polchinski84,wetterich93}, the 
fluctuations are progressively taken into account by introducing an infrared cutoff that enforces the decoupling of the low- and high-momentum 
modes at a running scale $k$. For $k=0$, all fluctuations are included and the exact theory is recovered. One ends up with flow equations for 
the moments of the renormalized disorder that describe the evolution of these moments as one decreases the infrared scale $k$. For instance, 
an equation is obtained for the second cumulant of the renormalized random field or random force $\Delta_k(q=0;\phi_1,\phi_2)$~\cite{fisher86b,narayan92,FRGledoussal-chauve,FRGledoussal-wiese,tarjus04,tissier06,tissier11}, which is the quantity already 
considered in the previous sections, with the field $\phi$ and  the infrared scale $k$ playing here the same role as the local magnetization 
$m$ and the inverse system size $1/L$, respectively. (This RG equation for $\Delta_k$ is in general part of a hierarchy of coupled flow 
equations.)

In order to reach the fixed point that controls the long-distance behavior under study, one must  introduce scaling dimensions and 
convert the quantities appearing in the RG flow equations from ``dimensionful'' to ``dimensionless''. For the cases of interest where 
on the one hand avalanches are present and on the other hand dimensional reduction is found in standard perturbation theory, we 
have stressed that the fixed point is at \textit{zero temperature}. Temperature is then a dangerously irrelevant variable, and an 
associated exponent $\theta >0$ is introduced through an appropriate definition of a renormalized temperature $T_k$~\cite{villainfisher1,villainfisher2,fisher86b,narayan92,FRGledoussal-chauve,FRGledoussal-wiese,tarjus04,tissier06,tissier11}: 
$T_k \sim k^\theta$. Near the zero-temperature fixed point, the dimension $d_{\phi}$ of the field $\phi$ is modified from its 
standard value of $(d-2+\eta)/2$, with $\eta$ the anomalous dimension, by a term involving the temperature exponent:
\begin{equation}
\label{eq_dim_phi}
d_{\phi} = \frac{1}{2}(d-2+\eta-\theta)=\frac{1}{2}(d-4+\bar \eta),
\end{equation}
where we have also introduced the additional anomalous dimension $\bar\eta$ through the relation $\theta=2+\eta-\bar \eta$. Similarly, 
the second cumulant of the renormalized random field or force has the scaling dimension of a $2$-point 1PI vertex, $2-\eta$, modified 
by the temperature exponent, \textit{i.e.} $2-\eta - \theta=-2\eta+\bar\eta$; it can be put in dimensionless form as
\begin{equation}
\label{eq_dim_delta}
\Delta_k(q=0;\phi_1,\phi_2)\sim k^{-(2\eta -\bar\eta)} \delta_k(0;\varphi_1,\varphi_2),
\end{equation}
where $\varphi$ is the dimensionless field (see also SI appendix).

Dimensional reduction corresponds to $\theta=2$, which implies $\bar\eta=\eta$, and to all other exponents equal to their value in the 
system without disorder in dimension $d-2$. The main outcome of the functional RG studies is that breakdown of dimensional reduction 
is related to the presence of a cusp in the functional dependence of the dimensionless second cumulant of the renormalized random 
field or force, $\delta_{k} (0;\varphi_1,\varphi_2)$, in the vicinity of the zero-temperature fixed point~\cite{fisher86b,narayan92,
FRGledoussal-chauve,FRGledoussal-wiese,tarjus04,tissier06,tissier11}. More concretely, after introducing 
$\varphi= (\varphi_1 +\varphi_2)/2$ and $\delta \varphi =(\varphi_2 - \varphi_1)/2 $, the ``cuspy'' behavior that changes the 
critical exponents from their dimensional reduction prediction is of the form
\begin{equation}
\label{eq_cusp_delta}
\delta_*(0; \varphi -\delta \varphi, \varphi+\delta \varphi) = \delta_{*,0}(\varphi) +  \delta_{*,a}(\varphi) \vert \delta \varphi \vert 
+ O(\delta \varphi^2)
\end{equation}
when $\delta \varphi \rightarrow 0$, with $\delta_{*,a}<0$; the star indicates the fixed-point value at $k=0$. As a result of a nonzero 
$\delta_{*,a}$, the exponent $\theta$ takes a nontrivial $d$-dependent value $<2$ and $\eta$ and $\bar\eta$ differ from the dimensional 
reduction values, with  $\bar\eta \neq \eta$.

The connection between the quantities computed through the functional RG and those discussed in the previous sections can be 
made by associating the infrared cutoff $k$ with the inverse of the linear extent of the system, \textit{i.e.} $k\sim 1/L$.  
Eq.~(\ref{eq_1PIfunction_cusp}) can then be expressed in a dimensionless form by dividing the cumulant $\Delta_L$ and the 
magnetization  $\delta m$ by their scaling dimensions $L^{2\eta-\bar\eta}$  and $L^{-(d-4+\bar\eta)/2}$, respectively. We 
immediately obtain that the amplitude of the linear cusp in dimensionless form scales as with
\begin{equation}
\label{eq_amplitude_cusp}
L^{d_f-2(2-\eta)-(2\eta-\bar\eta)-\frac{1}{2}(d-4+\bar\eta)}=L^{d_f-\frac{1}{2}(d+4-\bar\eta)}\; ,
\end{equation}
which can also be rewritten as $L^{d_f-(d-d_{\phi})}$. By comparison with Eq.~(\ref{eq_cusp_delta}), one therefore 
finds that the cusp persists in the dimensionless quantities when $L\rightarrow \infty$, \textit{i.e.} at the fixed point, if and 
only if $d_f=d-d_{\phi}$. If $d_f < d-d_{\phi}$, the cusp is only subdominant and does not affect the leading critical behavior 
and the associated exponents. (Note that the condition $d_f>d-d_{\phi}$ is not compatible with the result of the functional 
RG studies, in which proper renormalized theories have always been found with no stronger nonanalyticities than the 
linear cusp.)

We conclude from the above derivation that dimensional reduction breaks down due to avalanches if the fractal 
dimension of the largest typical ``critical'' avalanches satisfies the condition 
\begin{equation}
\begin{split}
\label{eq_condition_noDR}
d_f=d-d_{\phi}.
\end{split}
\end{equation}
On the other hand, dimensional reduction remains valid if
\begin{equation}
\begin{split}
\label{eq_condition_DR}
d_f<d-d_{\phi},
\end{split}
\end{equation}
despite the presence of the avalanches and of a cusp in the dimensionful  cumulants of the effective disorder at 
zero temperature. In the latter case, the difference $(d-d_{\phi}-d_f)$ can be reinterpreted and computed in the 
functional RG. Indeed when perturbing the ``cuspless'' fixed-point value\footnote[9]{Weaker nonanalyticities, \textit{i.e.} 
subcusps, could still be present but do not affect dimensional reduction~\cite{tarjus04,tissier06}).} of the dimensionless 
cumulant with a function that itself displays a linear cusp, the amplitude of the cuspy perturbation should go to zero 
as $k \rightarrow 0$ in such a way that
\begin{equation}
\label{eq_vanishing_cusp}
\delta_k(0;\varphi-\delta \varphi,\varphi+\delta \varphi)\simeq \delta_*(0;\varphi-\delta \varphi,\varphi+\delta \varphi) 
+k^{\lambda} f_{\lambda}(\varphi,\delta \varphi)
\end{equation}
with $\lambda=(d-d_{\phi})-d_f >0$ and $f_{\lambda}(\varphi,\delta \varphi) \simeq \vert \delta \varphi \vert f_{\lambda}(\varphi)$ 
when $\delta \varphi \rightarrow 0$. Information on the fractal dimension of the largest typical avalanches at criticality can then be obtained 
from an investigation of the irrelevant directions associated with nonanalytic eigenfunctions around the fixed point. This is 
what we have done by solving the nonperturbative functional RG equations derived in Ref.~\cite{tissier11} for a function 
$\delta_k$ of the form given in Eq.~(\ref{eq_vanishing_cusp}) (see SI appendix).

\textbf{Results and discussion.} 
We are now in a position to discuss the consequences of Eqs.~(\ref{eq_condition_noDR}) and (\ref{eq_condition_DR}) for several 
disordered systems in which dimensional reduction is predicted by standard perturbation theory.

Consider first the mean-field limit. Avalanches are present at $T=0$ and the distribution of the avalanche sizes at criticality can 
be described by the scaling expression in Eq.~(\ref{eq_distrib_avalanche_L}). The exponents $\tau$ and $d_f$ can be easily derived 
for fully connected models. This was first done by Dahmen and Sethna~\cite{dahmen96} for the out-of-equilibrium behavior of the 
slowly driven RFIM at zero temperature, but can be generalized to other models as well. (As usual, the mean-field exponents are 
expected to be  ``super-universal''.) The values of the avalanche exponents are $\tau=3/2$ and $d_f=4$. These values have also been 
recovered by Le Doussal and Wiese~\cite{dynamic_av-distrib_ledoussal} from a field-theoretical treatment of elastic manifolds in a 
random environment. What conclusion can then be drawn about the influence of the avalanches on the long-distance physics? At the 
upper critical dimension $d_{uc}$ at which the exponents take their mean-field values, the anomalous dimension $\bar\eta =0$ so that $d_{\phi}=(d_{uc}-4)/2$. The conditions in Eqs.~(\ref{eq_condition_noDR}) and (\ref{eq_condition_DR}) simply 
amount to comparing $d_f=4$ and $d_{uc}/2 +2$. For random field and random anisotropy models (we include here models with 
$N$-component fields which we expect to behave in a similar manner as that of the single-component one), the upper critical 
dimension is $d_{uc}=6$ so that dimensional reduction should apply. The same is true for the statistics of dilute branched polymer  
for which the upper critical dimension is $d_{uc}=8$~\cite{lubensky78}. On the other hand, for interfaces in a disordered environment, 
the upper critical dimension is $d_{uc}=4$: a failure of dimensional reduction is then expected, possibly  only in logarithmic 
corrections at $d=d_{uc}$ but more severe as one lowers the dimension.

As one decreases the dimension from the upper critical one, there must be a nonzero range of dimensions for which the dimensional 
reduction predictions correctly describe the critical behavior of the random field, random anisotropy and branched polymer models, 
but likely not that of the random manifold one. Actually, the latter model has been studied in great detail by the perturbative functional 
RG in $d=4-\epsilon$~\cite{static-distrib_ledoussal,dynamic_rosso,dynamic_av-distrib_ledoussal}. It was found that the dimension 
$d_f$ characterizing the cutoff on large avalanches in the presence of a finite-size or infrared cutoff on the system is equal to $d+\zeta$, 
where $\zeta$ is the exponent describing the roughness of the interface (for a single-component displacement field). As the dimension 
$d_{\phi}$ of the field is itself equal to $-\zeta$ ($\bar\eta$ is formally equal to $4-d-2\zeta$), it follows that the equality $d_f=d- d_{\phi}$ 
is always verified and that dimension reduction never applies, as indeed found by direct computation of the critical exponents within 
the functional RG or in computer simulations. This conclusion is valid for the pinned phase, in equilibrium, and for the depinning 
threshold in the driven case.

For the RFIM at equilibrium, we have shown through a nonperturbative
functional RG that dimensional reduction breaks down below a
nontrivial critical dimension $d_{cusp}\simeq
5.1$ (see Refs. \cite{tissier06,tissier11} and footnote 2). According to the above
conditions, the avalanche exponent $d_f$ should then be equal to
$d-d_{\phi}=(d+4-\bar\eta)/2$ below $d_{cusp}$ and to
$(d+4-\bar\eta)/2-\lambda$, where $\lambda$ is the eigenvalue
associated with the irrelevant cuspy directions around the cuspless
fixed point (see preceding section and SI appendix), above $d_{cusp}$. In Fig.~2, we plot
the theoretical prediction for $d_f$ based on the above relations and
the computation of $d_{\phi}$ and $\lambda$ from the solution of the
flow equations previously derived in our nonperturbative functional RG
approach of the RFIM~\cite{tissier11} (see also footnote 2). The prediction is confirmed at the upper (see
above) and lower critical dimensions. For the latter, $d_{lc}=2$, one
indeed expects the avalanches to be compact even at criticality and
their fractal dimension therefore to be equal to the spatial
dimension, $d_f=d=2$ (see also Ref.~\cite{perkovic99}).  As
the dimension of the field is equal to zero, $d_{\phi}=0$ (and
$\bar\eta=2$), the equality in Eq.~(\ref{eq_condition_noDR}) is
satisfied. (Note that the results at the lower and upper critical
dimensions apply to both the equilibrium and out-of-equilibrium
critical behavior of the RFIM\footnote[10]{It has also been suggested that the critical properties of the RFIM in and 
out of equilibrium are in the same universality class in all dimensions~\cite{perkovic99,perez04,liu09}}.) 
Beside this, direct measurements or computations of the avalanche exponent $d_f$ are unfortunately
scarce. We therefore suggest that, as done for models of an elastic
interface in a disordered environment~\cite{static_middleton,static-distrib_ledoussal,dynamic_rosso}
and for the out-of-equilibrium, metastable behavior of the driven
RFIM~\cite{perkovic99,sethna01,perez04,liu09}, systematic studies of
the avalanches and of the cumulants of the effective (renormalized)
disorder would be worthwhile, \textit{e.g.} in the ground state of the
RFIM, and would allow a direct test of the predictions made on the
basis of the functional RG.

Finally, for the statistics of dilute branched polymers, so long as dimensional reduction applies, $\bar \eta=2\eta$ and is negative. 
In  consequence, $d-d_{\phi}=d+4-\bar\eta >d+4$. As the fractal dimension $d_f$ should also be less than the dimension $d$ of 
space\footnote[11]{This reasoning does not apply to the elastic manifolds in a random environment. In this case, $d$ is the 
dimension of space on which the displacement field describing the location of the manifold is defined. The size of an avalanche, 
which represents the total amount by which the manifold moves, therefore involves both the ``lateral'' dimension (by an amount 
$\sim L^d$ for a system of lateral linear size $L$) and the change in the displacement field itself ($\sim L^{\zeta}$ with the 
roughness exponent $\zeta <1$). In this case, $d_f=d+\zeta < d+1$.}, one can see that $d_f\leq d<(d+4)/2$ when $d\lesssim 4$. 
From the condition on the scaling of the avalanches in Eq.~(\ref{eq_condition_DR}), we therefore obtain that dimensional reduction 
applies, at least, when $d\leq 4$ and in the vicinity of the upper critical dimension $d_{uc}=8$ (see above); the existence of an 
intermediate range of dimensions characterized by dimensional reduction breakdown is highly unlikely. This is of course in 
agreement with the known results according to which dimensional reduction (to the Lee-Yang edge singularity, or equivalently 
to the universal repulsive gas singularity, in two fewer dimensions) is always valid in the case of dilute branched 
polymers~\cite{brydges03,cardy03}.

\begin{figure}[ht]
\centering
\includegraphics[width=.9 \linewidth]{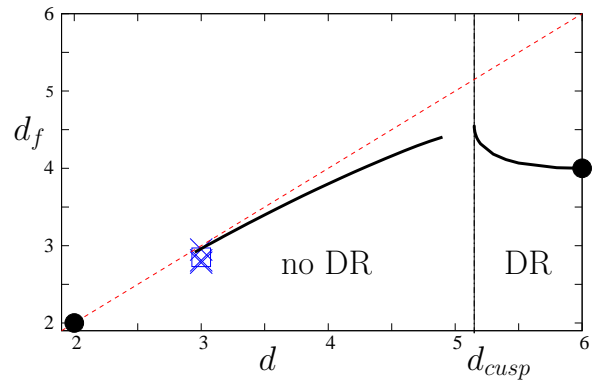}
\caption{ Fractal dimension $d_f$ of the largest typical avalanches versus space dimension $d$ for the RFIM at the equilibrium critical point,  as predicted by Eq.~(\ref{eq_condition_noDR}) (in the region ``noDR'') and Eq.~(\ref{eq_condition_DR}) (in the region  ``DR'') and the nonperturbative functional RG. The numerical resolution of the RG flow equations becomes extremely difficult in low dimension, typically for $d\lesssim 2.9$ and when approaching 
$d_{DR}\simeq d_{cusp}\simeq 5.15$ so that we have no result for these ranges of $d$. The filled circles indicate 
the known values at the lower and upper critical dimensions. The crosses are the numerical estimates for the out-of-equilibrium 
critical behavior of the driven RFIM in $d=3$~\cite{perkovic99,perez04,liu09} and the square that for the equilibrium behavior~\cite{liu09}. The dashed line is the upper bound ($d_f\leq d$).}
\end{figure}

To summarize, we have related the breakdown of  ``dimensional reduction''  to the scaling characteristics of ``avalanches''. 
The former is the formal property according to which the critical behavior in the presence of quenched disorder is the same 
as that of the clean system in two dimensions less and  is found within conventional perturbation theory in models 
whose long-distance physics is controlled by a zero-temperature fixed point, whereas the latter are  large-scale physical events 
taking place in the relevant configuration of the system (ground state in equilibrium or metastable state out of equilibrium) 
under the variation of an external source. This provides a solution to the puzzle of why dimensional reduction breaks down 
in some models and not in others, in some range of dimensions and not for others. 

Note finally that at small but nonzero temperature, there are \textit{no} avalanches and the variation of the relevant configuration 
of the system under a change of the external source is continuous, except possibly in mean-field models. The nonanalyticities in the 
functional dependence of the cumulants of the renormalized disorder and of the associated Green's functions are then rounded 
in ``thermal boundary layers''~\cite{BLchauve00,BLbalents04,BLledoussal10,tissier06}. In systems at equilibrium, these 
boundary layers are linked to the presence of low-energy excitations that may also take place on large scales at and near 
criticality and are described as ``droplets''~\cite{droplet_bray,droplet_fisher}. The relation between droplets and avalanches 
in disordered systems is by itself a very interesting topic, which however we have not considered here. In any case, this 
underscores that properties such as dimensional reduction and its breakdown or avalanches crucially depend on the system 
being at zero temperature or having its critical behavior controlled by a zero-temperature fixed point.

%% == end of paper:

%% Optional Materials and Methods Section
%% The Materials and Methods section header will be added automatically.

%% Enter any subheads and the Materials and Methods text below.
%\begin{materials}
% Materials text
%\end{materials}

%% Optional Appendix or Appendices
%% \appendix Appendix text...
%% or, for appendix with title, use square brackets:
\appendix[Nonperturbative functional RG for the RFIM]

We summarize here the main features of the nonperturbative functional RG description of the equilibrium critical behavior of RFIM 
developed in Refs.~\cite{tarjus04,tissier06,tissier11} as well as its extension to compute the stability of the zero-temperature fixed point 
against nonanalytic perturbations.

The central quantity is the so-called ``effective average action'' $\Gamma_k$~\cite{wetterich93,berges02} in which only 
fluctuations of modes with momentum larger than an infrared cutoff $k$ are effectively taken into account.  In the language 
of  magnetic systems, $\Gamma_k$ is the Gibbs free-energy functional of the local order parameter field obtained after a coarse-graining down to the (momentum) scale $k$. The effective average action obeys an exact RG equation under the variation of
the infrared cutoff $k$~\cite{wetterich93,berges02}.

In the presence of disorder, the generating or free-energy functionals are sample-dependent, \textit{i.e.} random, 
and should therefore be characterized either by their probability distribution or by their cumulants. The latter 
description is more convenient as it focuses on quantities, cumulants and associated Green's functions, which are translationally 
invariant and can be generated through the introduction of copies 
(or ``replicas'') of the original system that are submitted to distinct external sources~\cite{tarjus04,tissier11}. 
The effective average action that generates the cumulants of the renormalized disorder, $\Gamma_k[\{\phi_a\}]$, 
then depends on the local order parameter fields associated with the various copies $a$. It satisfies the 
following exact functional RG flow equation~\cite{tarjus04,tissier11}: 
\begin{equation}
\begin{aligned}
\label{eq_erg}
&\partial_k\Gamma_k\left[\{ \phi_a\}\right ]=\\&
\dfrac{1}{2} \int \frac{d^{d} q}{2\pi^d} \sum_{ab}  \partial_k R_{k}^{ab}(q^2) \bigg (\left[ \Gamma _k^{(2)}+ R_k\right]^{-1}\bigg)_{q,-  q}^{ab},
\end{aligned}
\end{equation}
where $\Gamma_k^{(2)}$ is the matrix formed by the second functional derivatives of $\Gamma_k$ with respect 
to the fields $\phi_a( q)$ and  $R_{k}^{ab}(q^2)=\widehat{R}_k(q^2) \delta_{ab}+ \widetilde{R}_k(q^2)$, where $\widehat{R}_k$ 
and $\widetilde{R}_k$ are infrared cutoff functions that enforce the decoupling of the low- and high-momentum modes at the 
scale $k$ (see below). Note that the formalism can be upgraded to a superfield theory in order to describe the physics directly at zero 
temperature (\textit{i.e.}, at equilibrium, the ground-state properties)~\cite{tissier11}: this allows one to make the 
underlying supersymmetry of the model explicit~\cite{parisi79} and to consider features associated with spontaneous 
or explicit breaking of the latter~\cite{tissier11}.

From Eq.~(\ref{eq_erg}), one can derive a hierarchy of coupled RG flow equations for the cumulants of the 
renormalized disorder, $\Gamma_{k1}[\phi_1]$, $\Gamma_{k2}[\phi_1,\phi_2]$, etc, that are obtained from 
$\Gamma_k[\{\phi_a\}]$ through an expansion in increasing number of unrestricted  sums over copies:
\begin{equation}
\label{eq_free_replica_sums}
\Gamma_k\left[\{ \phi_a\}\right ]=
\sum_a \Gamma_{k1}[\phi_a]-\frac{1}{2}\sum_{a,b}\Gamma_{k2}[\phi_a,\phi_b] + \cdots
\end{equation}
At the microscopic scale, say $k=\Lambda$, the effective average action reduces to the ``bare'' action (or effective 
hamiltonian) of the multy-copy system
\begin{equation}
\begin{aligned}
\label{eq_bare_action}
\Gamma_{k=\Lambda}\left[\{ \phi_a\}\right ]&= \int d^d x \sum_{a}\bigg\{\frac{1}{2}[ \partial  \phi_a( x)] ^2
+\frac{\tau}{2}   \phi_a( x) ^2\\&+ \frac u{4!}  \phi_a( x)^4 \bigg\}-\frac {\Delta_B}{2} \int d^d x
\sum_{a,b}   \phi_a(x) \phi_b( x),
\end{aligned}
\end{equation}
which generates the cumulants of the renormalized disorder at the mean-field level ($\Delta_B$ is the bare variance 
of the random field that is taken with a Gaussian distribution of zero mean). At the end of the flow, when $k=0$, all 
fluctuations are incorporated and one recovers the effective action $\Gamma[\{ \phi_a\} ]$ which is the generating 
functional of the  cumulants of the renormalized disorder and corresponds in the language of 
magnetic systems to the exact Gibbs free-energy functional of the multi-copy system.

The detour via the superfield formalism provides  a nonperturbative approximation scheme for the exact RG equation, Eq.~(\ref{eq_erg}), and a relation between the cutoff functions $\widehat{R}_k(q^2)$ and $\widetilde{R}_k(q^2)$ 
which, both, do not explicitly break the underlying supersymmetry at the origin of dimensional reduction. The minimal 
truncation of $\Gamma_k$ that  already contains the key features for a nonperturbative study of the long-distance 
physics of the RFIM is the following:
\begin{equation}
\label{eq_ansatz_Gamma_k}
\begin{split}
\Gamma_k\left[\{ \phi_a\}\right ]= & \int  d^d x  \sum_{a} \bigg\{ \frac 1{2} Z_{k}(\phi_a(x))  [\partial  \phi_a( x)]^2 
+  U_{k}( \phi_a( x)) \bigg\}\\&-\frac 1{2} \int  d^d x \sum_{a,b} V_{k}( \phi_a( x),  \phi_b( x)) ,
\end{split}
\end{equation}
with three functions $Z_k$, $U_k$ and $V_k$ to be determined. On the other hand, the cutoff functions must satisfy the relation 
$\widetilde{R}_k(q^2)=-(\Delta_k/Z_k)\partial_{q^2}\widehat{R}_k(q^2)$,
with $\Delta_k$ the strength of the renormalized random field and $Z_k$ the field renormalization constant. Inserting 
the above ansatz for $\Gamma_k\left[\{ \phi_a\}\right ]$ into Eq.~(\ref{eq_erg}) leads to a set of coupled flow equations
for the three functions $Z_{k}(\phi)$, $U_k(\phi)$ and $V_{k}( \phi_1,  \phi_2)$ (or alternatively its second derivative 
$\Delta_{k}( \phi_1,  \phi_2)=\partial_{\phi_1}\partial_{\phi_2}V_{k}( \phi_1,  \phi_2)$ which is the second cumulant 
of the renormalized random field at zero momentum discussed in the text). The RG is \textit{functional} as its central objects are functions 
instead of coupling constants.

One more step is needed to cast the nonperturbative functional RG flow equations in a form that is suitable for searching for 
the anticipated zero-temperature fixed points describing the critical behavior of the RFIM. One has to introduce appropriate scaling dimensions. 
This requires to define a renormalized temperature $T_k$ which should flow to zero as $k \rightarrow 0$. (This is the precise meaning of 
a ``zero-temperature'' fixed point.) Near such a  fixed point, one has the following scaling dimensions: 
\begin{equation}
T_k \sim k^\theta,\;  Z_{k} \sim k^{-\eta}, \; \phi_a  \sim k^{\frac{1}{2}(d-4+\bar \eta)},
\end{equation}
with $\theta$ and $\bar \eta$ related through $\theta=2+\eta-\bar \eta$, as well as
\begin{equation}
U_k\sim k^{d-\theta}, \;V_k \sim k^{d-2\theta},
\end{equation}
so that the second cumulant of the renormalized random field $\Delta_k$ scales as $k^{-(2\eta- \bar \eta)}$.

Letting the dimensionless counterparts of $U_k, V_k,\Delta _k,  \phi$  be denoted by lower-case letters, 
$u_k, v_k,\delta _k,  \varphi$,  the resulting flow equations can be symbolically written as
\begin{equation}
\label{eq_flow_dimensionless}
\begin{split}
&\partial_t u'_k(\varphi)=\beta_{u'}(\varphi),\\&
\partial_t z_k(\varphi)=\beta_{z}(\varphi),\\&
\partial_t \delta_k(\varphi_1,\varphi_2)=\beta_{\delta}(\varphi_1,\varphi_2),
\end{split}
\end{equation}
where $t=\log(k/\Lambda)$. The beta functions themselves depend on $u_k'$, $z_k$, $\delta_k$ and their derivatives [in addition, 
the running anomalous dimensions $\eta_k$ and $\bar\eta_k$ are fixed by the conditions $z_k(0)=\delta_k(0,0)=1$]. Their expressions are given in Ref.~\cite{tissier11}.

Fixed points are studied by setting the left-hand sides of the equations in Eq. (\ref{eq_flow_dimensionless} to zero. The zero-temperature 
fixed point controlling the critical behavior of the RFIM is once unstable and has been determined in a previous investigation~\cite{tissier11}. 
We found that above a dimension close to $5.15$, there exists a fixed point with no cusp singularity in the functional dependence of the 
associated $\delta_*(\varphi_1,\varphi_2)$. After introducing $\varphi= (\varphi_1 +\varphi_2)/2$ and $\delta \varphi =(\varphi_2 - \varphi_1)/2 $, 
the dimensionless second cumulant of the renormalized random field can indeed be expanded as
\begin{equation}
\label{eq_cuspless_delta}
\delta_*(\varphi -\delta \varphi, \varphi+\delta \varphi) = \delta_{*,0}(\varphi) + \frac{1}{2} \delta_{*,2}(\varphi)\delta\varphi^2  
+ O(\vert \delta \varphi \vert^3)
\end{equation}
when $\delta \varphi \rightarrow 0$. It can be shown that the critical behavior  then satisfies dimensional reduction: $\bar\eta=\eta$, so that 
$\theta=2$, and the critical exponents are exactly given by those of the pure Ising model in two dimensions less within the nonperturbative 
approximation that is the counterpart of Eq.~(\ref{eq_ansatz_Gamma_k}).

To compute the eigenvalue $\lambda$ that characterizes the stability of the above cuspless fixed point with respect to a ``cuspy'' perturbation, 
we have considered the vicinity of the fixed point with $u'_k$ and $z_k$ set at their fixed-point values and 
$\delta_k(\varphi-\delta \varphi,\varphi+\delta \varphi)\simeq \delta_*(\varphi-\delta \varphi,\varphi+\delta \varphi) 
+k^{\lambda} f_{\lambda}(\varphi,\delta \varphi)$ with $f_{\lambda}(\varphi,\delta \varphi) \simeq \vert \delta \varphi \vert f_{\lambda}(\varphi)$ 
when $\delta \varphi \rightarrow 0$. By linearizing the flow equation for $\delta_k$ around $\delta_*$ and expanding around 
$\delta \varphi= 0$ it  is easy to derive that $ f_{\lambda}(\varphi)$ satisfies the following eigenvalue equation:
\begin{equation}
 \label{eq_lambda}
\begin{split}
&\lambda f_{\lambda}(\varphi) = \frac{1}{2} (d - 4  +3 \eta ) f_{\lambda}(\varphi)  +  \frac{1}{2}  (d -4 + \eta ) \varphi f_{\lambda}'(\varphi) 
+ \\& v_{d}\,   \tilde{\partial}_{t} \int^{\infty}_{0} dy \: y^{\frac{d}{2}-1}  
\bigg\lbrace \frac{3}{2}  f_{\lambda}(\varphi) \Big(  4 z_{*}'(\varphi) p_{*}(y,\varphi) p^{(0,1)}_{*}(y,\varphi) +\\&
 4 [z_{*}(\varphi) + s' (y)]  p^{(0,1)}_{*}(y,\varphi)^{2}+[z_{*}''(\varphi) - \delta_{*, 2} (\varphi)]  p_{*}(y,\varphi)^{2}  \Big) +
\\&   3     f_{\lambda}'(\varphi)  p_{*}(y,\varphi)  \Big( 2[z_{*}(\varphi) + s' (y)]  p^{(0,1)}_{*}(y,\varphi)  
+ z_{*}'(\varphi) p_{*}(y,\varphi) \Big)
\\&+    f_{\lambda}''(\varphi)  [z_{*}(\varphi)+ s' (y)] p_{*}(y,\varphi)^2   \bigg\rbrace    ,
\end{split}
\end{equation}
where  $v_d^{-1}=2^{d+1}\pi^{d/2} \Gamma(d/2)$, derivatives of functions of a single argument are denoted by primes and partial derivatives 
are denoted by superscripts in parentheses; $y$ is the square of the dimensionless momentum, $s(y)$ is the (dimensionless) cutoff function 
defined from $\hat{R}_{k}(q^2) = Z_{k} k^2 s(q^2/k^2)$, and $p_{*}(y,\varphi ) = [ y z_{*}(\varphi) + s(y) + u_{*}''(\varphi) ]^{-1}$ is 
the (dimensionless) ``propagator'', \textit{i.e.} the $1$-copy, $2$-point Green's function. Finally, 
$\widetilde \partial_t$ is an operator acting only on the cutoff function $s(y)$ (appearing explicitly or through the dimensionless propagator) 
with $\widetilde \partial_t s(y) \equiv (2-\eta)s(y)-2ys'(y)$. (Choices of appropriate functional forms for $s(y)$ are discussed in \cite{tissier11}.) 
In deriving the above equation, we have used the fact that $\bar\eta=\eta$ and $\delta_{*,0}(\varphi)=z_{*}(\varphi)$,  which are  properties 
of the dimensional-reduction fixed point.

An equation for the fixed-point function $\delta_{*, 2} (\varphi)$ is also derived by inserting the expansion in powers of 
$\delta \varphi$ of $\delta_*$ [see Eq.~(\ref{eq_cuspless_delta})] in the corresponding beta function in Eq.~(\ref{eq_flow_dimensionless}). 
The algebra is straightforward but cumbersome and is not worth presenting here. 
From the knowledge of $u'_{*}(\varphi)$ and $z_{*}(\varphi)$, which are obtained from two coupled equations (see Ref.~\cite{tissier11}), 
we first solve the equation for $\delta_{*, 2} (\varphi)$  and then 
use the  input to solve  Eq.~(\ref{eq_lambda}). All partial differential equations are numerically integrated on a grid by 
discretizing the field $\varphi$. The corresponding ``cuspless'' fixed point only exists above a dimension $d_{DR}\simeq 5.15$. Note 
also that at the upper critical dimension,  $d_{uc}=6$,  one can analytically determine the solution of the above equations: 
as expected the fixed point is Gaussian and the eigenvalue $\lambda=1$.

The result for  $\lambda$ versus spatial dimension $d$ is displayed in Fig. \ref{fig_appendix}. Within numerical accuracy, the dimension $d_{cusp}$ at 
which $\lambda \rightarrow 0$ is indistinguishable from $d_{DR}$, namely $d_{cusp}\simeq d_{DR} \simeq 5.15$ (see footnote 2). 
We have used these values of $\lambda$ to construct the curve for the fractal dimension of the largest typical avalanches in Fig. 2.

\begin{figure}[ht]
\label{fig_appendix}
\centering
\includegraphics[width=.9 \linewidth]{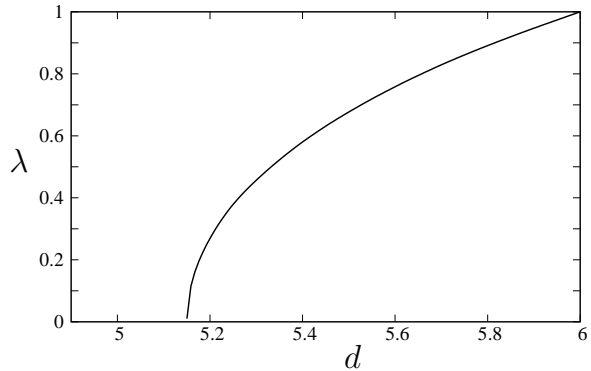}
\caption{ Variation with spatial dimension $d$ of the eigenvalue $\lambda$ associated with a ``cuspy'' perturbation 
around the ``cuspless'' fixed point corresponding to dimensional reduction. Below $d_{cusp}\simeq d_{DR} \simeq 5.15$, only ``cuspy'' 
fixed points are possible.}
\end{figure}
%% PNAS does not support submission of supporting .tex files such as BibTeX.
%% Instead all references must be included in the article .tex document. 
%% If you currently use BibTeX, your bibliography is formed because the 
%% command \verb+\bibliography{}+ brings the <filename>.bbl file into your
%% .tex document. To conform to PNAS requirements, copy the reference listings
%% from your .bbl file and add them to the article .tex file, using the
%% bibliography environment described above.  

%%  Contact pnas@nas.edu if you need assistance with your
%%  bibliography.

% Sample bibliography item in PNAS format:
%% \bibitem{in-text reference} comma-separated author names up to 5,
%% for more than 5 authors use first author last name et al. (year published)
%% article title  {\it Journal Name} volume #: start page-end page.
%% ie,
% \bibitem{Neuhaus} Neuhaus J-M, Sitcher L, Meins F, Jr, Boller T (1991) 
% A short C-terminal sequence is necessary and sufficient for the
% targeting of chitinases to the plant vacuole. 
% {\it Proc Natl Acad Sci USA} 88:10362-10366.

%% Enter the largest bibliography number in the facing curly brackets
%% following \begin{thebibliography}

\maketitle

\end{document}